\documentclass[aps,prd, nofootinbib,twocolumn]{revtex4-1}
\usepackage{amsmath,amsfonts,amssymb,graphicx,graphics,color, hyperref}
\usepackage[latin9]{inputenc}
\usepackage{pifont}
\usepackage{subfigure,epsfig,epstopdf}
\usepackage{bm}
\usepackage{enumerate}

\newcommand{\be}{\begin{equation}}
\newcommand{\ee}{\end{equation}}
\newcommand{\bea}{\begin{eqnarray}}
\newcommand{\eea}{\end{eqnarray}}

\begin{document}
 \def\be{\begin{equation}}
 \def\ee{\end{equation}}
 \def\l{\lambda}
 \def\a{\alpha}
 \def\b{\beta}
 \def\g{\gamma}
 \def\d{\delta}
 \def\e{\epsilon}
 \def\m{\mu}
 \def\n{\nu}
 \def\t{\tau}
 \def\p{\partial}
 \def\s{\sigma}
 \def\r{\rho}
 \def\sl{\ds}
 \def\ds#1{#1\kern-1ex\hbox{/}}
 \def\sla{\raise.15ex\hbox{$/$}\kern-.57em}
 \def\nn{\nonumber}
 \newcommand{\bth}{{\bf 3}}
 \newcommand{\btw}{{\bf 2}}
 \newcommand{\bon}{{\bf 1}}
 \def\Tr{\textnormal{Tr}}
 \def\th{\theta}
 \def\({\left(}
 \def\){\right)}
 \def\[{\left[}
 \def\]{\right]}

\newcommand{\Eq}[1]{Eq.~(\ref{#1})}
\newcommand{\Fig}[1]{Fig.~\ref{#1}}
\newcommand{\Sec}[1]{Sec.~(\ref{#1})}

\preprint{}

\title{Equation of state of imbalanced cold matter  from chiral perturbation theory}
\author{Stefano Carignano}
\email{stefano.carignano@lngs.infn.it}
\author{Andrea Mammarella}
\email{andrea.mammarella@lngs.infn.it}
\author{Massimo Mannarelli}
\email{massimo@lngs.infn.it}

\affiliation{INFN, Laboratori Nazionali del Gran Sasso, Via G. Acitelli, 22, I-67100 Assergi (AQ), Italy}

\begin{abstract}
We study the thermodynamic properties of matter at vanishing temperature for non-extreme values of the isospin chemical potential and of the strange quark chemical potential. From the leading order  pressure  obtained by maximizing the static chiral Lagrangian density we derive a simple expression for the equation of state in the pion condensed phase and in the kaon condensed phase.   We find an analytical expression for the maximum of the  ratio between the energy density and the Stefan-Boltzmann energy density as well as  for the isospin  chemical potential at the peak both in good agreement with lattice simulations of quantum chromodynamics.  We speculate on the location of the crossover from the Bose-Einstein condensate state to the Bardeen-Cooper-Schrieffer  state by a simple analysis of the thermodynamic properties of the system.  For  $\mu_I \gtrsim 2 m_\pi$ the leading order chiral perturbation theory  breaks down;   as an example it  underestimates  the energy density of the system and leads to a wrong asymptotic behavior.  \end{abstract}

\maketitle
\section{Introduction}
Nuclear matter at sufficiently high density can be described as  an imbalanced system having  non-vanishing isospin density and, in some circumstances, strangeness. The most  remarkable and extreme  example is matter in the  interior of compact stars, which  is believed to be  neutron-rich  with a small proton fraction~\cite{Shapiro-Teukolsky}. Eventually,  in the core of compact stars  hyperonic nucleation can take place, leading to a non-vanishing strangeness density. In any case, understanding the properties of matter at nonzero isospin and strangeness density allows us to explore quantum chromodynamics (QCD) in a regime in which various methods can be employed with significant overlap. 

The isospin chemical potential, $\mu_I$, and the strange quark chemical potential, $\mu_S$, have several effects  on matter. Confining ourselves to the mesonic sector, the obvious effect is  a  Zeeman-like mass splitting  within multiplets. A nontrivial effect  is a rotation of  the vacuum in flavor space. Indeed, sufficiently large values of  $\mu_I$, and/or of $\mu_S$, are able to tilt the chiral condensate leading  to  meson condensation. The grand canonical phase diagram of matter at vanishing temperature as a function  of  $\mu_I$ and  $\mu_S$ has been firmly established by chiral perturbation theory ($\chi$PT), see~\cite{Kogut:2001id}. Varying $\mu_I$ and $\mu_S$, three different phases can be realized: the normal phase, the pion condensed ($\pi c$) phase and the kaon condensed ($Kc$) phase. 

Very powerful methods for the analysis of imbalanced quark matter   have been developed in lattice QCD (LQCD)~\cite{Alford:1998sd, Kogut:2002zg, Detmold:2012wc, Detmold:2008yn,Cea:2012ev, Endrodi:2014lja}. Remarkably, LQCD simulations are feasible at nonzero $\mu_I$ and $\mu_S$. Unfortunately these simulations are not easy to perform  for physical meson masses and/or large  $\mu_I$ and $\mu_S$.   Moreover, the interpretation of the LQCD results requires a careful understanding of the physics, which is certainly more transparent when using effective field theories or  perturbative QCD (pQCD).
 The meson condensed phases have  also been studied by Nambu-Jona Lasinio (NJL) models in~\cite{Toublan:2003tt, Barducci:2004tt, Barducci:2004nc, Ebert:2005wr,Ebert:2005cs, He:2005nk, He:2010nb} and by random matrix models in~\cite{Klein:2004hv, Kanazawa:2014lga}.  In~\cite{Barducci:2004tt, He:2005nk} the $\mu_I$-$T$  phase diagram was obtained by  NJL models. Finite temperature effects were also considered in  $\chi$PT~\cite{Loewe:2002tw, Loewe:2004mu, Xia:2014bla}. In particular, in~\cite{Loewe:2002tw,  Loewe:2004mu} the $\mu_I$-$T$  phase diagram has been obtained. 

The thermodynamic properties of the meson condensed phases have been studied by LQCD simulations in~\cite{Detmold:2012wc, Detmold:2008yn}. Previously, various  results on  the $\pi c$ phase were derived   in~\cite{He:2005nk} by an NJL model, including an expression of   the equation of state (EoS). Recently in~\cite{Andersen:2015eoa,Cohen:2015soa, Graf:2015pyl} a perturbative analysis of QCD at large isospin density has been presented. Those pQCD  results are consistent with LQCD  for $\mu_I \gtrsim 3 m_\pi$~\cite{Graf:2015pyl}, where $m_\pi$ is the pion mass. At smaller values of $\mu_I$ it seems that pQCD  underestimates the energy density and is not able to capture the condensation mechanism. However, for small values of  $\mu_I$ (and $\mu_S$), $\chi$PT can be used.  Although the $\chi$PT isospin density of the system has been determined several years ago~\cite{Son:2000xc}, to the best of our knowledge a careful study of the EoS of imbalanced matter has never been derived within this framework.

By the present manuscript we fill this gap, analyzing the thermodynamic properties of cold matter both in the $\pi$c phase and in the $Kc$ phase at non-extreme values of $\mu_I$ and/or $\mu_S$. We    use a realization of  $\chi$PT that includes only the pseudoscalar mesons~\cite{Gasser:1983yg, Leutwyler:1993iq, Ecker:1994gg,  Scherer:2002tk, Scherer:2005ri}, thus our results are valid for $|\mu_I| < 770$ MeV, $|\mu_S| < 550$ MeV and for baryonic chemical potentials below the nucleon mass, see for example the discussion in~\cite{Kogut:2001id} and in~\cite{Mammarella:2015pxa}. We derive a remarkable simple expression for the leading order (LO)  EoS. As we will see, the comparison with the LQCD results of~\cite{Detmold:2012wc} will clarify some interesting aspects of the energy density in the $\pi c$ phase. Moreover,  we estimate the  value of $\mu_I$ corresponding to the  crossover from the Bose-Einstein condensate (BEC) state  to the Bardeen-Cooper-Schrieffer (BCS) state. 

When comparing our results with those obtained by different methods  we use consistent values of $m_\pi$ and of the pion decay constant, $f_\pi$.  In particular, the  LQCD simulations of~\cite{Detmold:2012wc} and the pQCD results of~\cite{Graf:2015pyl} are obtained for $m_\pi = 390 $ MeV. Given the rather large value of the pion mass,  we take $f_\pi =110$ MeV, see for example~\cite{Colangelo:2001df}.  

The present paper is organized as follows. In Sec.~\ref{sec:mesoncond}, we briefly review the meson condensed phases. In Sec.~\ref{sec:EoS}, we derive the expression of the equation of state and compare our result for the energy density with that obtained by LQCD simulations and by pQCD. In Sec.~\ref{sec:conclusions}, we draw our conclusions. 

\section{Meson condensation}\label{sec:mesoncond}
 
The  LO  Lagrangian density describing the  in-medium pseudoscalar mesons  can be written as \cite{Gasser:1983yg}
\be\label{eq:Lagrangian_general}
{\cal L} = \frac{f_\pi^2}{4} \text{Tr} (D_\nu \Sigma D^\nu \Sigma^\dagger) + \frac{f_\pi^2}{4} \text{Tr} (X \Sigma^\dagger + \Sigma X^\dagger )\,,
\ee
where $\Sigma$ corresponds to the meson fields, $X=2 B M$ with $M= \text{diag}(m,m,m_s)$ and the trace is in flavor space.   The pion and the kaon masses can be expressed in terms of the quark masses   by the usual relations $m_\pi^2=2 B m $ and 
$m_K^2=B (m+m_s)$, see~\cite{Gasser:1983yg, Leutwyler:1993iq, Ecker:1994gg,  Scherer:2002tk, Scherer:2005ri}.
The in-medium effects can be introduced by means of external currents in the covariant derivative~\cite{Gasser:1983yg}. For the case of interest we define 
\be
D_\mu \Sigma = \partial_\mu\Sigma - \frac{i}{2} [v_\mu,\Sigma]\,, \qquad v_\mu=-2 \mu \delta_{\m 0} \,,
\ee
 with\footnote{We do not include the baryonic chemical potential because mesons have zero baryonic charge.}  $\mu=\text{diag}(\mu_I/2,-\mu_I/2,-\mu_S)$. Since we focus on the zero-temperature thermodynamics, we neglect fluctuations and replace $\Sigma \to \bar \Sigma$, 
where $\bar \Sigma$ is the vacuum expectation value of the mesonic fields. In the normal phase  $\bar\Sigma=I$, but a generic vacuum  depends on $N_f^2-1$ parameters, corresponding to the possible orientations in $SU(N_f)_V$ space. However,  it can be shown that many vacua are degenerate, see~\cite{Kogut:2001id} and the discussion in~\cite{Mammarella:2015pxa}. The $N_f=2$ vacuum Lagrangian depends only on one angle, $\alpha$, and the three-flavor vacuum on two angles, $\alpha$ and $\beta$.  As in~\cite{Kogut:2001id} we consider the three-flavor  parametrization
\begin{widetext}
\begin{eqnarray}
\bar \Sigma=\left( \begin{array}{ccc} 1 & 0 & 0 \\
                                 0 & \cos \beta & -\sin \beta \\
                                 0 & \sin \beta & \cos \beta
                   \end{array} \right) \;
            \left( \begin{array}{ccc} \cos \alpha & \sin \alpha  & 0 \\
                                 -\sin \alpha & \cos \alpha & 0 \\
                                 0 & 0 & 1
                   \end{array} \right) \;
            \left( \begin{array}{ccc} 1 & 0 & 0 \\
                                 0 & \cos \beta & \sin \beta \\
                                 0 & -\sin \beta & \cos \beta
                   \end{array} \right),
\label{SadP}
\end{eqnarray}
\end{widetext}
with $\alpha,\beta \in (0,\pi/2)$. The explicit form of the ground state is found maximizing the static Lagrangian
\be
\mathcal{L}_{0}=-\frac{f_\pi^2}{4} \Tr\[\mu, \bar{\Sigma} \] \[\mu,\bar{\Sigma}^\dag \]+\frac{f_\pi^2 B}{2} \Tr\[M (\bar{\Sigma}+\bar{\Sigma}^\dag)\]\,, \label{eq:L02}
\ee
with respect to the angles $\alpha$ and $\beta$. The maximum of the static Lagrangian corresponds to the the tree-level pressure of the system at vanishing temperature.

In this picture the isospin and strange quark chemical potentials can lead to an increase of the pressure if they 
are able to tilt the vacuum in a direction in $SU(N_f)_V$ space. When this happens, the system undergoes a phase transition to a meson condensed phase.
More in detail, by varying $\mu_I$ and $\mu_S$ it is possible to show that three different phases can be realized: the normal phase (which at zero temperature corresponds to the vacuum), the $\pi c$ phase and the $K c$ phase, characterized by a non-vanishing pion and kaon condensate, respectively~\cite{Kogut:2001id}. 
We will normalize the pressure of the vacuum to zero by subtracting to the tree-level results
  the pressure $p^0 \equiv \mathcal{L}_{0}^N$ corresponding to the maximum of the static Lagrangian in the normal phase.
  
The normal phase is stable for $ \mu_I <m_\pi  \text{ and }  \mu_S <m_K-\frac12 \mu_I\,$. In this case
$\bar{\Sigma}_N=\text{diag}(1,1,1)\, $ and the vacuum pressure is given by 
$p^0_2=f_\pi^2 m_\pi^2$  for  $N_f=2$ and
$p^0_3=f_\pi^2 m_\pi^2 \left(\frac{1}2 + \frac{m_K^2}{m_\pi^2}\right)$ for  $N_f=3$.

The $\pi c$ phase is stable for $ \mu_I>m_\pi$ and  $\mu_S < (-m_\pi^2+\sqrt{(m_\pi^2-\mu_I^2)^2+4 m_K^2 \mu_I^2})/(2 \mu_I)$, and is characterized by
$\cos \alpha_{\pi}=m_\pi^2/\mu_I^2$ and  $\beta_{\pi} =0\,$. In this phase  the LO normalized pressure is given by~\cite{Son:2000xc,Kogut:2001id} \be
p^{\pi c}_\text{LO} = \frac{f_\pi^2\mu_I^2}{2} \left(1- \frac{m_\pi^2}{\mu_I^2}\right)^2 \,, \label{eq:pressure-pions}
\ee
which does not depend on the kaon mass and on the strange quark chemical potential. The reason is that this pressure is determined by the condensation of pions, and therefore it can only depend on the properties of pions. This expression is valid for both $N_f=2$ and  $N_f=3$. 

The $K c$ phase is stable for 
$
\mu_I>2(m_K- \mu_S)$ and $
\mu_S> (-m_\pi^2+\sqrt{(m_\pi^2-\mu_I^2)^2+4 m_K^2 \mu_I^2})/(2 \mu_I)$, 
and is characterized by
$
\cos \alpha_K=\left( \frac{m_K}{\frac12 \mu_I+\mu_S}\right)^2$ and $ \beta_K=\pi/2\,$. The LO normalized pressure  
is analogous to Eq.~\eqref{eq:pressure-pions} with the replacement $m_\pi \to m_K$ and $\mu_I\to \mu_K =\mu_I/2 +\mu_S$, related to the fact that kaons have isospin $1/2$ and strangeness $1$.

\section{Equation of state}\label{sec:EoS} 

The  energy density of the system can be obtained by
\be
\epsilon=  \mu_I n_I + \mu_S n_S - p\,,
\label{eq:energy_density_general}
\ee
where $n_I$ and $n_S$ are the isospin and strange number densities, respectively. 
Given that the considered  phases  are characterized by the condensation of pions or of kaons, we can rewrite the energy density as follows
\be
\epsilon^{\pi c}=  \mu_I n^{\pi c}_I - p^{\pi c}\,,
\label{eq:energy_density_pic}
\ee
for the $\pi c$ phase (since $n^{\pi c}_S = 0$), and 
\be
\epsilon^{Kc}=  \mu_K n^{Kc}_K - p^{Kc}\,,
\label{eq:energy_density_pic_k}
\ee
for the $K c$ phase. The LO isospin number density in the $\pi c$ phase can be calculated from
\be
n_{I, \text{LO}}^{\pi C}=\frac{\partial p^{\pi C}_\text{LO}}{\partial \mu_I}=f_\pi^2 \mu_I \(1-\frac{m_\pi^4}{\mu_I^4}\)\,,
\label{eq:nI}
\ee
which agrees with the result obtained in~\cite{Son:2000xc}. The kaon number density in the $K c$ phase  has an analogous expression, with $m_\pi \to m_K$ and $\mu_I \to \mu_K$, see~\cite{Detmold:2008yn}.

\begin{figure}[t!]
\includegraphics[width=.45\textwidth]{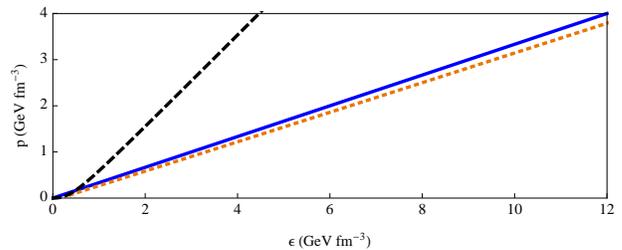}
\caption{(color online)  The  equation of state obtained with $\chi$PT (dashed black line) compared with the ideal gas case (solid blue line) and with the pQCD result of~\cite{Graf:2015pyl} (dotted orange line). }
\label{fig:EoS-pions}
\end{figure}

Upon substituting Eq.~\eqref{eq:nI} in Eq.~\eqref{eq:energy_density_pic} we obtain the LO energy density in the $\pi$c phase 
\be
\epsilon^{\pi c}_\text{LO}=\frac{f_\pi^2  \mu_I^2}{2}\left(1+2\frac{ m_\pi^2}{\mu_I^2}  - 3 \frac{m_\pi^4}{\mu_I^4}\right)\,.
\label{eq:epsilon-pions}
\ee
Inverting Eq.~\eqref{eq:pressure-pions}  we get $\mu_I(p)$, that allows us to obtain the EoS \be
\epsilon^{\pi c}_\text{LO}(p) = 2\sqrt{p(2f_\pi^2 m_\pi^2+p)}-p \label{eq_state}\,,
\ee
which is an increasing function of $p$ and vanishes for $p=0$. 
The expression of the energy density in the $K c$ phase can be  obtained from Eq.~\eqref{eq_state} with the replacement $m_\pi \to m_K$. 
In Fig.~\ref{fig:EoS-pions} we report a plot of the EoS for the $\pi c$ phase and we compare the result with the pQCD findings of~\cite{Graf:2015pyl} and with the ideal gas case.  From this figure it is clear that the agreement with pQCD is rather poor. It is interesting to note that  the conformal relation $\epsilon=3 p$ is satisfied for \be
\bar\mu_I = \sqrt{3} m_\pi\,,\label{eq:conformal}\ee corresponding as well to the point at which the energy density convexity changes. This coincidence is suggestive of a possible change from two different regimes. Since the conformal limit separates the  BEC state  from the BCS state, see~\cite{ketterle-review, Nishida:2010tm} for  a non-relativistic  discussion, it is tempting to interpret $\bar\mu_I $ as the chemical potential corresponding to the BEC-BCS crossover. Although our result is consistent with the NJL findings of~\cite{He:2010nb}, a more careful analysis  is necessary for  substantiating this conjecture.

A comparison of the LO expression of the pressure and energy with 
the NJL results has been done~\cite{He:2005nk}. In agreement with~\cite{He:2005nk} we obtain that 
the  LO $\chi$PT    expression of the energy density and of the pressure are consistent with  the corresponding NJL quantities for $\mu_I \lesssim 2 m_\pi$.

\begin{figure}[t!]
\includegraphics[width=.4\textwidth]{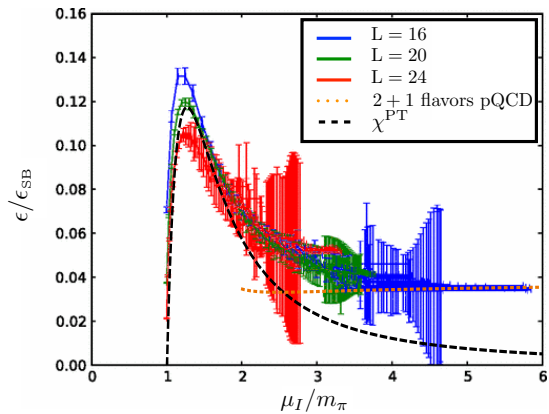}
\caption{(color online) Energy density over the Stefan-Boltzmann limit. The lattice points have been obtained at $T=20$ MeV; the  colors correspond to the different lattice volumes considered in~\cite{Detmold:2012wc}. The orange dotted line corresponds to the pQCD results of~\cite{Graf:2015pyl}. The dashed black line corresponds to our results.  }
\label{fig:energy_density}
\end{figure}

For comparing  the $\chi$PT energy density  with the results obtained by LQCD simulations in~\cite{Detmold:2012wc} and by pQCD in~\cite{Graf:2015pyl},  we divide it by the Stefan-Boltzmann limit, which has been defined in~\cite{Detmold:2012wc}   as $ \epsilon_{SB}=9 \m_I^4/(4 \pi^2) $. In Fig.~\ref{fig:energy_density} we report our ratio $\epsilon^{\pi c}_\text{LO}/\epsilon_\text{SB}$ and the results of~\cite{Detmold:2012wc} and~\cite{Graf:2015pyl}. We immediately notice that the $\chi$PT curve perfectly captures the peak structure at low $\mu_I$, while it begins to depart from the LQCD  results after $\m_I \sim 2 m_{\pi}$, indicating the breakdown of the LO approximation. An interesting result is that within our framework we can obtain an analytic expression for the position of the peak in this ratio, which for the $\pi c$ phase is given by 
\be
\label{eq:mupeak}
\mu_I^\text{peak} = \({\sqrt{13} -2}\, \)^{1/2} m_\pi \simeq 1.276\, m_\pi \,,
\ee
and is independent of $f_\pi$. This result is very close to the LQCD results obtained in~\cite{Detmold:2012wc},    where the values $ \mu_I^\text{peak} = \{1.20,1.25,1.275\} m_\pi  $  have been obtained considering  different spatial volumes $L^3$ with side   $L=\{16,20,24\}$, respectively. The  continuum-linearly-extrapolated value for the peak position 
 is $ \mu_I^\text{peak} = 1.30\,m_\pi$. 

\begin{figure}[t!]
\includegraphics[width=.4\textwidth]{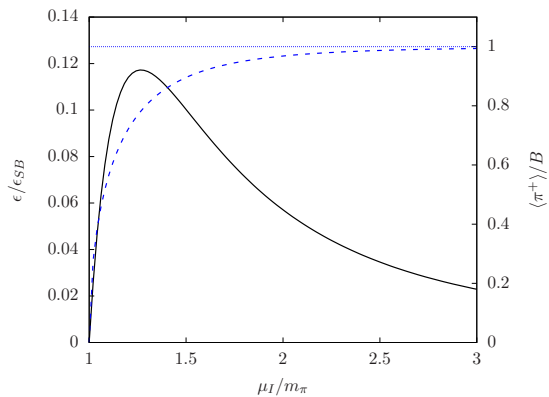}
\caption{(Color online) $\chi$PT results for the  ratio $\epsilon/\e_{SB}$ (solid black line) and the pion condensate normalized to its maximum value $B$ (dashed blue line). }
\label{fig:pioncond}
\end{figure}

In~\cite{Detmold:2012wc} the top of the peak is interpreted as the point
where the pion condensation sets in. However, in our case the pion
condensate has already reached a significant value at that point, almost
approaching its asymptotic limit $B$, see
Fig.~\ref{fig:pioncond}.
More specifically, the pion condensate can be evaluated by introducing external source terms 
in the chiral Lagrangian and by differentiating with respect to them. After the phase transition it grows as 
$\langle \pi^+ \rangle = B \sqrt{1-{m_\pi^4}/{\mu_I^4 }}$~\cite{Splittorff:2000mm, Kogut:2000ek,Kogut:2001id}, 
so that at $\mu_I^{\text{peak}}$ it reaches already $\sim 79\%$ of its asymptotic value.
 We are therefore more inclined to interpret this peak structure as a consequence of the filling of the condensate, rather
than of the onset of its formation, which at zero-temperature
should occur  at $\mu_I = m_\pi$.  We also obtain an analytic expression for the ratio at the peak
\be
\left.\frac{\epsilon}{\epsilon_{SB}}\right\vert_\text{peak} =\frac{4 (\sqrt{13}-5) \pi^2 }{9(-2+\sqrt{13})}\frac{f_\pi^2}{m_\pi^2}\,,
\label{eq:enpeak}
\ee
which would give information on the $f_\pi/m_\pi$ scaling if precise LQCD data  were available.
It is  worth mentioning that the above results hold in the $Kc$ phase if one considers 
$ \epsilon_{SB}=9 \mu_K^4/(4 \pi^2) $. Then, we obtain $\mu_K^\text{peak} = (\sqrt{13} -2)^{1/2} m_K$  and an expression analogous to Eq.\eqref{eq:enpeak}, with $m_\pi \to m_K$.

From Fig.~\ref{fig:energy_density} it seems that for $\mu_I > 2 m_\pi$ the LO $\chi$PT breaks down, resulting in an underestimate of  the energy density. The basic reason is that for large $\mu_I$ we have from Eq.~\eqref{eq:epsilon-pions} that $\epsilon^{\pi c}_\text{LO} \propto f_\pi^2 \mu_I^2$, while pQCD correctly predicts  $\epsilon^{\pi c} \propto  \mu_I^4$.     At the  next-to-leading order (NLO), it is possible to show that the $\chi$PT energy density  includes a term proportional to  $(2 l_1+2l_2 + l_3) \mu_I^4$, where $l_1$, $l_2$ and $l_3$ are low energy constants~\cite{Scherer:2002tk}.  Comparing our results with the pQCD energy density we obtain \be\label{eq:LEC} 2 l_1+2l_2 + l_3 = \frac{3}{8\pi^2}\left(\frac{\epsilon_\text{pQCD}}{\epsilon_\text{SB}} - \frac{2 f_\pi^2 \pi^2 }{9\mu_I^2}  + {\cal O}(\mu_I^{-4})\right)  \,,\ee leading to $2 l_1+2l_2 + l_3 \simeq0.6 \times 10^{-3}$ for $\mu_I= 3 m_\pi$, consistent with the empirical values~\cite{Scherer:2002tk}.

\section{Conclusions}\label{sec:conclusions}
We have derived an analytic expression for the EoS of cold mesonic matter in the $\pi c$ phase, reported in Eq.~\eqref{eq_state}. The LO $\chi$PT results agree remarkably well with LQCD simulations  and with the NJL results  for $\mu_I \lesssim 2 m_\pi$, confirming that LO $\chi$PT is able to capture the relevant low-energy physics  of the system. This is a delightful result given the extremely simple structure of the LO chiral  Lagrangian density. 
The analytic expression for the ratio of the LO energy density to the Stefan-Boltzmann limit at the peak and for the value of the chemical potential at the peak, reported respectively in  Eq.~\eqref{eq:enpeak} and in Eq.~\eqref{eq:mupeak} are in good agreement with the LQCD results, paving the way for a more quantitative comparison between $\chi$PT and LQCD, if precise lattice data  will be available. The  behavior of the LQCD data for $\mu_I \gtrsim 3 m_\pi$ can be used to constrain some NLO constants, as in Eq.~\eqref{eq:LEC}. The results obtained in the $\pi c$ phase can be easily extended to the  $K c$  phase. 

We have   observed that the energy density to pressure ratio reaches the conformal limit at $\bar \mu_I$ given in Eq.~\eqref{eq:conformal}, which corresponds as well to the point of vanishing curvature of the energy density, suggestive of a BEC-BCS crossover.  Whether  the $\chi$PT can capture some aspects of the BEC-BCS crossover is a tantalizing possibility, which should be further explored with  a detailed analysis of the system, see for example~\cite{Brauner:2008td}. 

There are several directions for improving the present work. One possibility is to systematically analyze   all NLO  temperature and  chemical potential corrections for a precise matching with LQCD data. Ultimately,  a realistic description of dense matter in compact stars will require the inclusion of baryonic degrees of freedom,  for example by means of the heavy-baryon effective theory, see~\cite{Scherer:2002tk}.

\begin{acknowledgments}
We would like to thank   J.~Schaffner-Bielich and W.~Detmold for  discussion. 
\end{acknowledgments}


%

\end{document}